\documentclass[11pt,twoside]{article}

\usepackage{asp2014}

\aspSuppressVolSlug
\resetcounters

\begin{document}

\title{Managing target of opportunity (ToO) observations at Observatorio Astrof{\'i}sico de Javalambre (OAJ)}

\author{Tamara~Civera$^1$}
\affil{$^1$Centro de Estudios de F{\'i}sica del Cosmos de Arag{\'o}n (CEFCA), Teruel, Arag{\'o}n, Spain; \email{tcivera@cefca.es}}

\paperauthor{Tamara~Civera}{tcivera@cefca.es}{0000-0002-0358-8503}{Centro de Estudios de F{\'i}sica del Cosmos de Arag{\'o}n (CEFCA)}{DPAD}{Teruel}{Arag{\'o}n}{44001}{Spain}



\begin{abstract}
The Observatorio Astrof{\'i}sico de Javalambre (OAJ) is a Spanish astronomical ICTS (Unique Scientific and Technical Infrastructures) located in the Sierra de Javalambre in Teruel (Spain). It has been particularly conceived for carrying out large-sky multi-filter surveys. As an ICTS, the OAJ offers Open Time to the astronomical community, offering more than 25\% through Legacy Surveys, Regular Programs (RP) and Director discretionary time (DDT). Regarding the RP, a new call for proposals is made public each semester accepting only proposals under the modality of Target of Opportunity (ToO). 

This contribution summarizes how ToOs are managed at OAJ presenting the different applications designed and implemented at the observatory to deal with them: the Proposal Preparation portal (to request observing time), the Phase2 Observing tool and the submitphase2 web service (to trigger the ToOs), the TAC Tracking portal (for telescope operators to support the observations) and the TACData portal (to publish and offer the images and their data products).
\end{abstract}



\section{Introduction}

The Observatorio Astrof{\'i}sico de Javalambre (OAJ\footnote{\url{https://oajweb.cefca.es/}}) is a Spanish astronomical ICTS facility (Unique Scientific and Technical Infrastructures) located in the Sierra de Javalambre in Teruel (Spain). It has been particularly conceived for carrying out large-sky multi-filter surveys and consists of two main telescopes of large field of view (FoV): JST250, a 2.5m with 3.4 square degrees FoV and JAST80 with 2 square degreess FoV. The Unit for Data Processing and Archiving (UPAD), as the needed data center to deal with the images from both telescopes, is also considered an essential infrastructure of the ICTS.

As an ICTS, the OAJ is committed to offer at least 20\% of the observing time. It actually offers more than 25\% of Open Time to the astronomical community through Legacy Surveys, Regular Programs (RP) and Director discretionary time (DDT). Regarding the RP, a new call for proposals is made public each semester accepting only proposals under the modality of Target of Opportunity (ToO). These projects are defined as proposals for which the target and/or observation epoch is not known at proposal submission. Moreover, depending on the scientific requirements and merit they could require a fast response after a trigger and they could be granted with ``override'' status (i.e. can interrupt the execution of another observation in the queue). Another aspect is that they may need a quick access to the data products and, in those cases, these are provided in a matter of minutes after the observations are made.

\section{Requesting observing time: Proposal Preparation portal}

Two calls for RP proposals are open each year: from 1st August to 15th September for semester A and from 1st February to 15th March for semester B. A web portal, the Proposal Preparation portal \footnote{\url{https://oajweb.cefca.es/}} (figure \ref{ex_fig1}) has been implemented to allow users to create proposals to apply for ToO time. This portal has functionalities like letting to edit the proposals until the deadline or automatically calculating the total requested time. This total requested time includes overheads and is based on different parameters like the time per exposure, the number of exposures and the number of triggers defined in the proposal.

When the call finished, the OAJ Time Allocation Committe (OAJ-TAC), a panel currently constituted by 7 members, evaluates the semester proposals. The Proposal Preparation portal also provides tools for the OAJ-TAC to visualize, evaluate the proposals and communicate the decision (which can be approved, partially approved or rejected) to the research team of the ToO proposal.

\articlefigure[width=.70\textwidth]{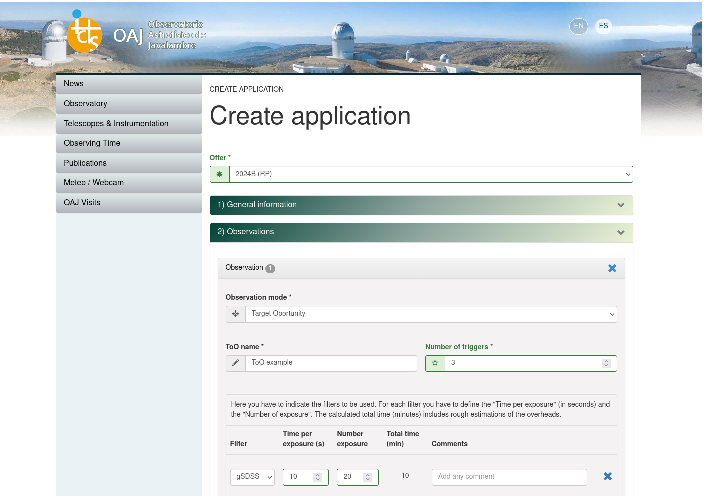}{ex_fig1}{Proposal Preparation portal: user interface to create a proposal}

\section{Triggering the ToO: Phase2 Observing tool and submitphase2 web service}

Once the proposal has been approved and the semester starts, users can trigger a ToO observation using the Phase2 Observing tool desktop application or the submit web service. Both tools allow the user to define and submit the coordinates of the target and the associated Observing Block (OB) properties like the number of nights it is observable. The Phase 2 Observing tool offers a user-friendly interface while the submitphase2 web service is intended to trigger them in an automated way using JSON format to describe them.

ToOs can be triggered by any authorized member of the research team of the ToO proposal until the number of total triggers and the total allowed time have not been reached. If required, a project global user may be provided to automatically trigger the ToOs via the submitphase2 web service.

\section{Receiving the ToOs at OAJ: TAC Tracking portal}

When a ToO is triggered, it is immediately sent to OAJ and automatically loaded in the TAC Tracking portal (figure \ref{ex_fig2}). This web portal has been implemented to help the telescope operators on the efficient planning of the observations at any given time along the night; it also permits checking the status of the observations of the different projects; and actually command the selected OBs to the JAST80 telescope through the creation of CSV files following a specific format (\citet{10.1117/12.2313208}). The different observations status operators can manage are: 'not observed' (the pointing must be observed), 'CSV generated' (the command file for the telescope has been generated and sent to the telescope), 'observed to be validated' (the pointing has been observed, but images must be still checked) and 'observed' (the pointing has been observed and images checked and validated).

Moreover, depending on the scientific requirements a fast response could be needed after a trigger that, depending on its merit, it could be granted with the ``override'' status over other existing proposals. In all cases, the trigger is announced through an alert system.

\articlefigure[width=.75\textwidth]{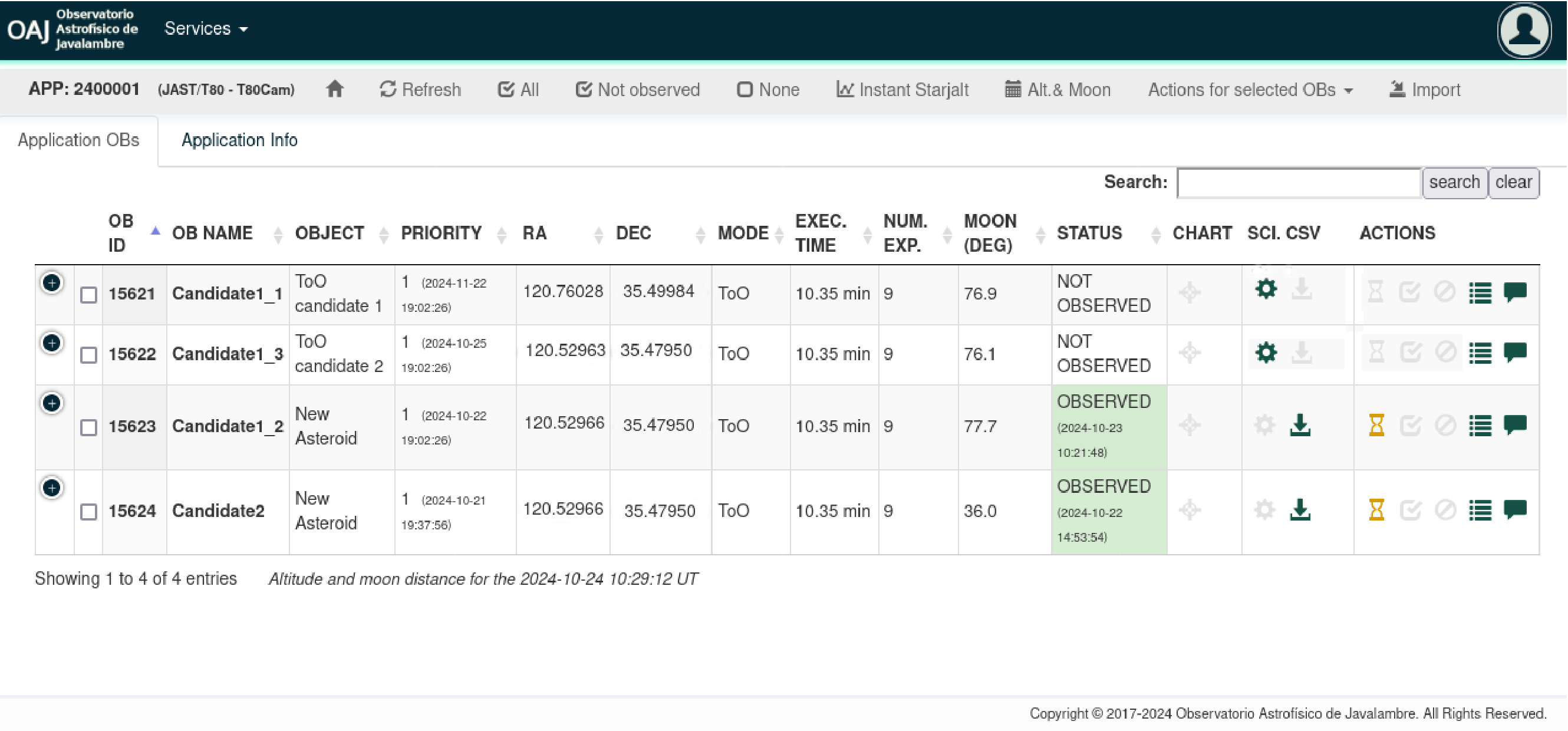}{ex_fig2}{TAC Tracking portal: user interface used by telescope operators to support ToO observations}

\section{Data publication: TACData portal}

When the ToO is observed and validated, its images are stored, processed and calibrated in a standard way with the most-recent OAJ pipelines in the Unit for Data Processing and Archiving (UPAD) (\citet{2019A&A...622A.176C}). Raw scientific data, calibration frames and the scientific data reduced and calibrated which includes images and data products (catalogs,...) are offered through the TACData portal \footnote{\url{https://tacdata.cefca.es/}} (figure \ref{ex_fig3}). This portal has been implemented to allow the research team of the ToO proposal to perform searches and to download the data and files. To make file downloading easier, the portal also includes functionality to create scripts to download a set of files using command line tools. Moreover, an email is sent to them when new project data is offered through the portal. The ToO research team may need a quick access to the data products and, in those cases, these are provided in a matter of minutes after the observations are made.

\articlefigure[width=.70\textwidth]{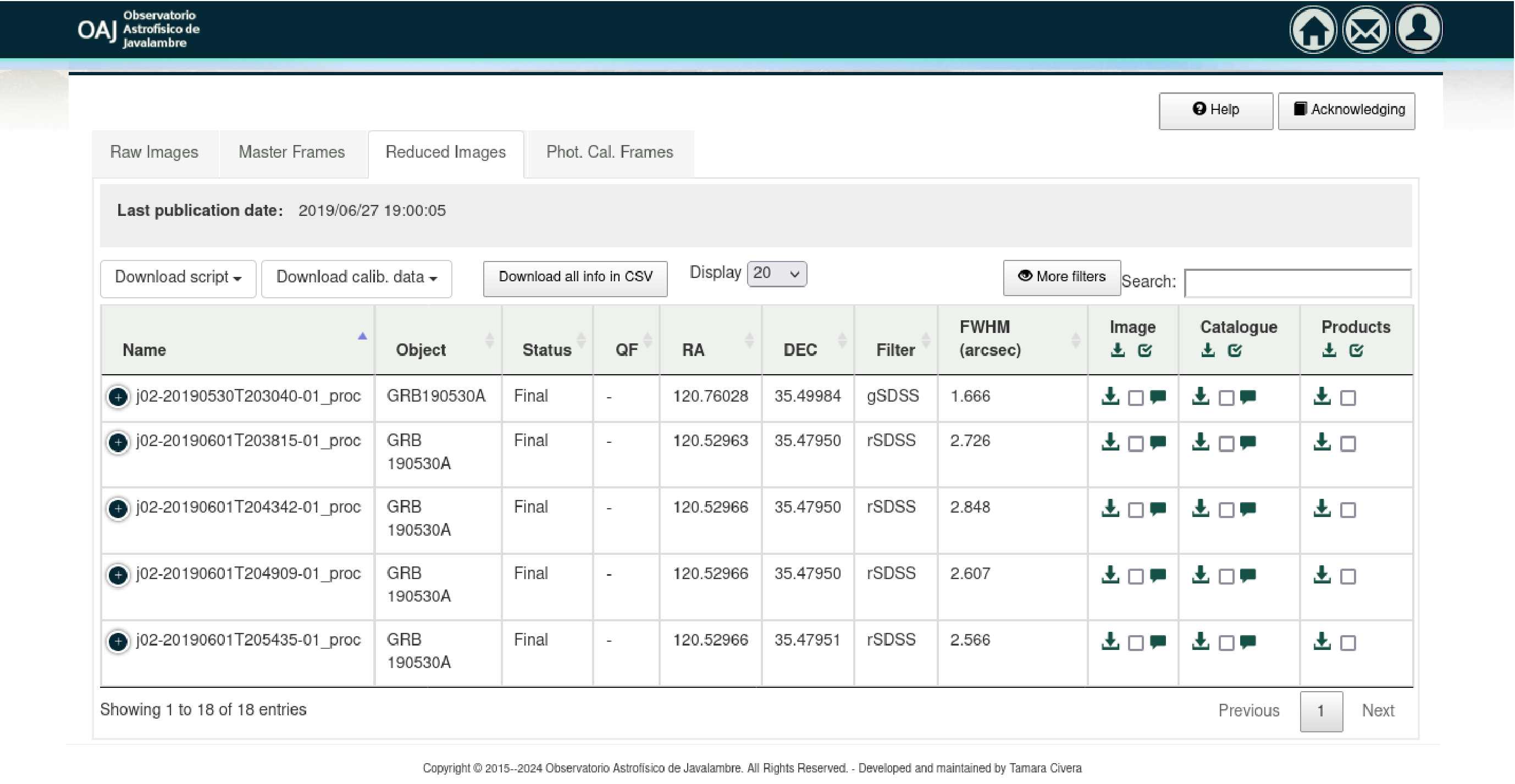}{ex_fig3}{TACData portal: user interface to search and download scientific data reduced}

\section{Conclusions}

The way ToOs are managed at OAJ has been summarized here by describing the different applications designed and implemented at the observatory to deal with them. In particular, the Proposal Preparation portal (to request observing time), the Phase2 Observing tool and the submitphase2 web service (to trigger the ToOs), the TAC Tracking portal (for telescope operators to support the observations) and the TACData portal (to publish and offer the images and their data products) have been presented.

\acknowledgements Funding for CEFCA has been provided by the Governments of Spain and Arag{\'o}n through the Fondo de Inversiones de Teruel; the Arag{\'o}n Government through the Research Groups E16\_20R; the Spanish Ministry of Science, Innovation and Universities (MCIN/AEI/10.13039/501100011033 and FEDER, \emph{Una manera de hacer Europa}) with grants PID2021-124918NB-C41, PID2021-124918NB-C41, PID2021-124918NB-C42, PID2021-124918NA-C43 and PID2021-124918NB-C44. 

\bibliography{P116}  


\end{document}